\documentclass[review]{elsarticle}

\usepackage{lineno,hyperref}
\modulolinenumbers[5]

\journal{Journal of \LaTeX\ Templates}

\begin{document}

\begin{frontmatter}

\title{A new approach to search for free neutron-antineutron oscillations using coherent neutron propagation in gas.}
%\tnotetext[mytitlenote]{Fully documented templates are available in the elsarticle package on %\href{http://www.ctan.org/tex-archive/macros/latex/contrib/elsarticle}{CTAN}.}

\author{V.\,Gudkov \corref{mycorrespondingauthor}}
\cortext[mycorrespondingauthor]{Corresponding author}
\ead{gudkov@sc.esu}
\address{Department of Physics and Astronomy, University of South Carolina, South Carolina, USA-29208}

\author{V.V.\,Nesvizhevsky}
\address{Institut Max von Laue -- Paul Langevin, 71 avenue des Martyrs, Grenoble, France-38042}

\author{K.V.\,Protasov}
\address{Laboratoire de Physique Subatomique et de Cosmologie, UGA-CNRS/IN2P3, Grenoble, France-38026}

\author{W.M.\,Snow}
\address{Department of Physics, Indiana University, 727 E. Third St., Bloomington, Indiana, USA-47405}

\author{A.Yu.\,Voronin}
\address{P.N. Lebedev Physical Institute, 53 Leninsky prospect, Moscow, Russia-119991}

\begin{abstract}
Coherent forward neutron propagation in gas is discussed as a new approach to search for neutron-antineutron oscillations ($ n-\bar{n}$), which violate both $B$ and $B-L$ conservation.  We show that one can increase the probability of neutron - antineutron transitions in the presence of a nonzero external magnetic field to essentially free neutron oscillation probability by tuning the density of an appropriate mixture of gases so that the neutron optical potential of the gas cancels that from the magnetic field.
\end{abstract}


\end{frontmatter}

%\linenumbers
\section{Introduction}

Neutron-antineutron ($n-\bar{n}$) oscillations would violate the conservation law so far observed for baryon number by two units. Sensitive searches for processes that violate the conservation of baryon number ($B$) such as proton decay and $n-\bar{n}$ oscillations and also processes that violate lepton number ($L$) conservation have long been of fundamental interest due to the many implications such a discovery would imply for particle physics and cosmology \cite{Stueckelberg38,Sakharov67,Kuzmin70,Georgi74,Zeldovich76,Hooft78,Davidson1979,MohapatraPRL80,Kazarnovskii80,Kuo80,Chang80,
MohapatraPLB80,Chetyrkin81,Dolgov81,Cowsik81,Rao82,Misra83,Rao84,Kuzmin85,Fukugita86,Shaposhnikov87,Dolgov92,Huber01,
Babu01,Nussinov02,Babu06,Dutta06,Berezhiani06,Bambi07,Babu09,Mohapatra09,Dolgov10,Gu11,Morrissey12,Arnold13,Canetti13,
Babu13,Gouvea14,Berezhiani16,Grojean18,Berezhiani2019,Shrock2019,Shrock_2_2019,Shrock_3_2019}. Cosmological arguments which use the Sakharov criteria \cite{Sakharov67} to generate the baryon asymmetry of the universe starting from a $B=0$ condition require $B$ violation. Many theoretical models possess $\Delta B=2$ processes leading to $n-\bar{n}$ without giving proton decay in the most popular $\Delta B=1$ channels \cite{Babu01,Nussinov02,Babu06,Dutta06,Babu09,Mohapatra09,Babu13,Arnold13,Dev2015,Berezhiani16,Dev2018}. A class of models called post-sphaleron baryogenesis (PSB) (see for example \cite{Babu13} and references therein) can generate the baryon asymmetry below the electroweak scale. $n-\bar{n}$ oscillation physics and the closely-related process of $n$ -- mirror $n$ oscillations has inspired several other recent investigations on a broad variety of relevant topics in both theory and experiment~\cite{Aharmim17, Dune14, HyperK18, Golubeva18, Barrow2019, Rinaldi2019, Babu15,  Mohapatra16, Kamyshkov17, Gardner18, Gardner15, Fujikawa16, McKeen16, Berezhiani15, Gardner16, Berezhiani2019, Berezhiani:2020nzn, BerezhianiIJMP04, Berezhiani:2008bc, Serebrov08, Serebrov09, Altarev09, BerezhianiEPJC12, BerezhianiPRD17}.

Experimental searches for $\Delta B=2$ processes involving neutrons have been conducted both by searching for antineutron appearance in a free neutron beam and through energy release in underground detectors from antineutron annihilation from oscillating neutrons bound inside nuclei \cite{Kuo80,Dover83,Dover89,Gal00,Chung02,Friedman08,Abe15}. The best free neutron oscillation searches have used a slow neutron beam passing through a magnetically-shielded vacuum chamber to a thin
annihilation target surrounded by a low-background antineutron annihilation detector. Antineutron annihilation in a target downstream of a free neutron beam is a spectacular experimental signature. An essentially background-free search is possible, and any positive signal can be extinguished by a very small change in the ambient magnetic field. The best constraint on $\tau_{n\to\bar{n}}$ with free $n$ used an intense cold neutron beam at the Institute Laue-Langevin (ILL) \cite{Baldo94} which built on earlier searches~\cite{Fidecaro85, Bressi90}.  The ILL experiment used a cold neutron beam from their 58 MW research reactor with a neutron current of 1.25$\times$10$^{11} {\it n}/{\rm s}$ incident on the annihilation target and achieved a limit of $\tau_{n-\bar{n}} > 0.86\times10^{8}$ $\rm{s}$~\cite{Baldo94}. The average velocity of the cold neutrons was $\sim$ 600 ${\rm m/s}$ and the average neutron observation time was $\sim$ 0.1 ${\rm s}$. A vacuum of $P\simeq 2\times10^{-4}$ ${\rm Pa}$ maintained in the neutron flight volume and a magnetic field of $|{\vec B}| < 10$ ${\rm nT}$ satisfied the quasi-free conditions for oscillations to occur~\cite{Bitter91, Kinkel94, Schmidt92}. Antineutron appearance was sought through annihilation with a $\sim$ 130 ${\rm \mu m}$ thick carbon film target which generated at least two tracks (one due to a charged particle) in the tracking detector with a total energy above 850 MeV in the surrounding calorimeter. In one year of operation the ILL experiment saw zero candidate events with zero background~\cite{Baldo94} using a tracking detector with several cm spatial resolution for the annihilation vertex.

The practical experimental figure of merit for a free neutron $n - \bar n$ search using this approach is $N_n t^2$, where $N_n$ is the total number of free neutrons observed in the experiment and $t$ is the observation time for free neutron propagation. An ambitious project at the European Spallation Source (ESS) \cite{Phillips15} proposes to increase the sensitivity by a factor $G\approx 10^2-10^3$ by using an advanced version of the ILL approach. It requires a dedicated beamline optimized for the production of slow neutrons with a large solid angle neutron extraction from the source to take full advantage of the improved phase space acceptance of supermirror neutron optics, which is the main enabling technology for the improvement in sensitivity by increasing $N_n$. The relatively large scale of the nontrivial single-bounce elliptical focusing supermirror, vacuum chamber, and magnetic shielding and the associated expense to realize this approach encourages thought on additional methods for increasing $N_n t^2$ and for economizing on the apparatus required.

\section{Standard experimental approach to the search for $n/\bar{n}$ oscillations}

Free neutron oscillation searches conducted to date have been designed so that the neutrons avoid interactions with matter and external fields. The motivation behind this strategy was to minimize the energy difference $\Delta E$ between the neutron and antineutron states during the observation time $t$. In practice even the best magnetic shielding still leaves a large enough residual magnetic field that $\Delta E \gg \varepsilon$, where $\varepsilon$ is the off-diagonal mixing term in the effective Hamiltonian for the $n/\bar{n}$ two-state system. Still the oscillation rate is not greatly suppressed if the ``quasi-free'' condition $(t\Delta E/\hbar)<1$ is met, where $\hbar$ is the reduced Planck constant. In this so-called ``quasifree" regime, the relative phase shift between the $n$ and $\bar{n}$ states, $e^{-i\Delta Et/\hbar}$, is small enough that the oscillation probability still grows quadratically with $t$ for short observation times and therefore the sensitivity of the measurement is not compromised.

We recently suggested \cite{Nesvizhevsky:2018uet,NGPSV,NGPSV2} that one could also conduct a sensitive $n-\bar{n}$ experiment in which one allows the freely propagating $n/\bar{n}$ of meV energies to reflect from $n/\bar{n}$ optical mirrors between the neutron source and the antineutron detector. We showed that the probability of coherent reflection of $n/\bar{n}$ from matter can be high and the relative phase shift can still be small enough to meet the quasifree condition in certain neutron beam phase space regimes (for an earlier analysis for ultracold neutrons see~\cite{Gol89, Yoshiki92}). The value of this observation lies in the additional flexibility that it can give for the optimization of the experiment with sufficient knowledge of the low energy antineutron-nucleus interaction as well as in increasing the experimental sensitivity or/and decreasing its cost. For slow neutrons the $\bar{n}$ coherent scattering amplitude comes from a single s-wave scattering length whose real and imaginary parts can be calculated within a phenomenological model~\cite{Batty01} reflecting a simple geometrical picture of $\bar{n}A$ annihilation. The strong $\bar{n}$ absorption on the nuclear surface means that the real part of the scattering amplitude is very close to the nuclear size plus the nuclear skin thickness, and the imaginary part of the scattering amplitude is approximately the same for all nuclei~\cite{Batty01,Karmanov:2000in}. This contrasts with the neutron case, where the absence of such strong absorption can lead to resonances whose effect on the low energy scattering amplitude can vary strongly for different isotopes. This relative simplicity of the low energy antineutron interaction with nuclei combined with the existing experimental data on antinucleon interactions can constrain the antineutron-nucleus optical potential to sufficient accuracy to allow an intelligent choice of the mirror material to be made for such an experiment.

\section{Description of the new proposed method}

In this work we point out another option for the realization of a free $n-\bar{n}$ experiment which exploits the coherence of antineutron forward scattering in a gas rather than the coherence of antineutron reflection from a mirror. We show that one can in principle operate a free $n-\bar{n}$ experiment in a magnetic field that can be much larger than employed in past searches if one also introduces gas in the neutron path whose pressure and composition is chosen so that the difference in the neutron and antineutron optical potential $V_{n, opt} -V_{\bar{n}, opt}$ in the gas cancels the difference  $2\mu B$ in the neutron and antineutron potential energies in the magnetic field well enough that the quasifree condition is still maintained. We show that such a choice is possible for one neutron polarization state and for  realistic values of the neutron and antineutron optical potentials in matter and that the additional attenuation of the neutron beam through the gas and the effects of the incoherent interactions of the neutrons with the gas can be acceptably small for the experiment. This idea shares some similarities with analogous ideas developed for the consideration of neutron-mirror neutron oscillations~\cite{Berezhiani2019}. Similar to our previous work, the value of this observation lies in the additional flexibility that it can give for the optimization of the experiment. In particular it makes it possible to imagine conducting the experiment in a much larger residual magnetic field than used in past searches provided it is sufficiently uniform, thereby relaxing one of the most severe technical requirements for the experiment. The practical implementation of this idea requires a degree of understanding of the low energy antineutron-nucleus interaction comparable in accuracy to that needed for the mirror reflection idea.

For low energy neutrons the coherence in forward neutron scattering through materials is very well established both by theory and experiment (see, for example~\cite{Rauch2015, Sears} and references therein), and the concept of the forward index of refraction is well known to operate as expected in the case of neutrino oscillations in matter through the MSW mechanism that theoretically explained the physical mechanism behind the solar neutrino deficit~\cite{MSW1,MSW2}. Still the coherence in scattering from a low density medium such as a gas for neutron-antineutron oscillations strikes many as nonintuitive. If  the scattering of the projectile from atoms in the gas is a simple classical two-body collision that   changes the energy and momentum of both the projectile and the target, decoherence is generated. This view, stated in some previous calculations of decoherence in oscillations occurring in a gas medium~\cite{Feinberg1961, Kerbikov2018},  is simply  incorrect (although the calculations presented later by these authors are fine). We refer to~\cite{CohScatt} for a clear discussion of the persistence of the coherent forward amplitude during propagation of a projectile through a gas of free atoms. The key point is that the projectile can avoid all the gas atoms (thereby transferring no energy or momentum to the atoms) and  remain coherent after passing through the gas sample while also accumulating the phase shifts which comprise the forward scattering amplitude.  The projectile-atom interactions which transfer energy and momentum to the individual atoms are the subset of events which give rise to the incoherence and generate diffuse scattering.  It is small impact parameters that contribute to the scattering cross section, and large impact parameters that contribute to the phase shift, and it is at large impact parameters, when scattering is avoided, that the phase shift is linear in the interaction strength. Since the cross section is quadratic in the interaction strength, the forward amplitude accumulates much faster than decoherent scattering can destroy it.

The arguments of this paper apply also to the low energy $s$-wave scattering of neutrons and antineutrons, and we can apply neutron optics theory to the propagation of antineutrons in gases as well. The value of the neutron index of refraction can be written as
\begin{equation}\label{index}
n^2=1+\frac{4\pi}{k^2}\sum_iN_if^i,
\end{equation}
where $N_i$ is the number of nuclei of type $i$ per unit volume, $k$ is the neutron wave number,  and $f^i$ is the neutron elastic forward scattering amplitude on an type-$i$ nucleus.
In our case, it is convenient to use the neutron Fermi potential which is directly related to the refractive index \cite{Gurevich,Squires,UCN}
\begin{equation}\label{fermif}
V=-\frac{2\pi \hbar^2}{m}\sum_iN_if^i,
\end{equation}
where $m$ is neutron mass.
For slow neutrons in the absence of resonances, the expression for the Fermi potential in terms of neutron coherent scattering
lengths $b_i$ becomes
\begin{equation}\label{fermib}
V=\frac{2\pi \hbar^2}{m}\sum_iN_ib^i.
\end{equation}

Let's use this Fermi potential to describe neutron - antineutron oscillations for neutron propagation in gases in the presence of a magnetic field $\vec{B}$. The mixing matrix for this case can be written as
\begin{equation}\label{mixMatr}
\cal{M}= \left(
\begin{array}{cc}
m_n -\vec{\mu_n} \cdot \vec{B} +V_n & \delta m \\
\delta m & m_n +\vec{\mu_n} \cdot \vec{B}+V_{\overline{n}} \\
\end{array}
\right) ,
\end{equation}
where $\delta m$ is a free neutron-antineutron mixing parameter, $\mu_n$ is neutron magnetic moment, and $V_n$ and $V_{\overline{n}}$ are Fermi potentials for neutron and antineutron, respectively.
The diagonalization of this matrix gives mass eigenstates related to pure neutron $|n>$ and antineutron $|\overline{n}>$ states
\begin{equation}\label{eigenst}
\left(
\begin{array}{c}
|n_1 > \\
|n_1 > \\
\end{array}
\right)
= \left(
\begin{array}{cc}
\cos \theta & \sin \theta \\
- \sin \theta & \cos \theta \\
\end{array}
\right)
\left(
\begin{array}{c}
|n> \\
|\overline{n}> \\
\end{array}
\right)
\end{equation}
with
\begin{equation}\label{tan}
\tan (2\theta) =\frac{2 \delta m}{(2\vec{\mu_n} \cdot \vec{B} -V_n+V_{\overline{n}})}.
\end{equation}
This leads to the probability to find an antineutron at time $t$ starting from an initial pure neutron state at time $t=0$ as
\begin{equation}\label{prob}
P_{n \overline{n}}(t)=\sin^2(2\theta )\sin^2(\Delta E t/2),
\end{equation}
where
\begin{equation}\label{delE}
\Delta E = \left[ (2\vec{\mu_n} \cdot \vec{B} -V_n+V_{\overline{n}})^2+4(\delta m)^2 \right]^{1/2}.
\end{equation}
We neglect the imaginary parts of the Fermi potentials in this expression for now. We will consider the corrections from neutron beam attenuation at the end of the paper.
Let us first estimate the difference of Fermi potentials $\Delta V=V_n-V_{\overline{n}} $, which depends on the values of neutron and antineutron the scattering lengths in eq.(\ref{fermib}). The neutron scattering lengths are well known (see, for example~\cite{Mughabghab} and references therein), and are usually dominated by contributions from potential scattering for slow neutrons. The absolute values of potential scattering lengths fluctuate around the value of the nuclear radius
\begin{equation}\label{bn}
b_{nA}=1.35 A^{1/3},
\end{equation}
for scattering on a nucleus with atomic number $A$, and most scattering lengths are positive. It should be noted that this behavior for the absolute values of $b_{nA}$ is in very good agreement with optical model calculations (see~\cite{PhysRev.96.448, Mughabghab} and references therein). What is more important for our case, it was shown~\cite{PhysRev.71.145} that increasing the value of the imaginary part of the optical potential, which corresponds to increasing the neutron absorption, leads to a smoother behavior of $b_{nA}$ around the value of nuclear radius with smaller fluctuations. This fact justifies a similar approximation for the theoretically predicted values of real parts of antineutron scattering lengths
\begin{equation}\label{bantin}
b_{\overline{n}A}=1.54 A^{1/3},
\end{equation}
which was used in~\cite{NGPSV} (imaginary parts appear to be close to 1 fm). Unfortunately, there are no experimental measurements of $b_{\overline{n}A}$.

The behavior of the real part of scattering length given in eq.(\ref{bantin}) was obtained in a
model describing antinucleon-nucleus annihilation by a complex potential with strong
imaginary part (see~\cite{Batty01} and references therein). This model predicts the
$A^{1/3}$-behavior and gives the same sign of the scattering length for $n-A$ and for $\overline{n}-A$ systems. This behavior of the scattering length is explained by an annihilation which
strongly suppresses the wave function inside the nucleus: it gives the real part proportional the nucleus size and the imaginary part proportional to nuclear surface diffuseness. An alternative approach based on chiral effective field theory proposed recently in~\cite{Dai:2017ont} could become a promising approach to determine the scattering length in the future.

The values of $b_{nA}$ for some gases are: for $^{16}O$ $b_{nO}=5.8 $ fm, for hydrogen $b_{nH}=-3.7 $ fm, and for $^{4}He$ $b_{nHe}=3.3 $ fm. This leads to an oxygen related neutron potential at atmospheric pressure of $V_n =7.5 \cdot 10^{-11} $ eV. For comparison, the value of the magnetic energy $|\vec{\mu_n} \cdot \vec{B}| =6 \cdot 10^{-17} $ eV for the magnetic field $B=10^{-9} T$.  Taking into account that the values $ V_n$ and $V_{\overline{n}} $ for a gas mixture are just the sums of corresponding optical potentials of the particular gases and that they are proportional to the density of these gases, one can set to zero the parameter
\begin{equation}\label{delCoh}
S=(2\vec{\mu_n} \cdot \vec{B} -V_n+V_{\overline{n}})
\end{equation}
that suppresses the oscillation rate in eqs.(\ref{tan}) and (\ref{delE}) by creating a gas mixture and applying the appropriate uniform magnetic field. This observation is the main result of this paper.
For example, choosing gases with a small slow neutron absorption like parahydrogen and $^{4}$He, which have opposite signs for neutron scattering lengths, we can zero the value of $S$,
which for $^4$He-H case is:
\begin{equation}\label{HeH}
S_{H-He}=(2\vec{\mu_n} \cdot \vec{B} -V_{nH}-V_{nHe}+V_{\overline{n}H}+V_{\overline{n}He})
\end{equation}
by adjusting three parameters: the density of parahydrogen, the density of helium, and the value of the residual magnetic field.
The choice of gases composed of light nuclei provides an easier target for the theoretical calculation of the antineutron scattering lengths. The antineutron scattering length calculation is also easier for nuclei like $^{12}$C and $^{16}$O, which opens up several additional choices for the gas mixture.

The statistical accuracy of the experiment compared to one with no gas is only slightly lowered as in the regime of gas densities and magnetic fields of interest for slow neutron $n-\overline{n}$ experiments the contribution from even the antineutron absorption on the nuclei of the gas atoms is small for the practical case of a slow neutron traveling 100 meters from the source to an antineutron detector. The neutron and antineutron scattering cross sections are smaller than the antineutron absorption cross sections, and the incoherent scattering from the gas at finite temperature would make a small additional contribution to the background in the antineutron detector from neutron capture gamma rays in the apparatus walls  compared to other sources. Note that this absorption does place a practical upper bound to the gas density. As there is a theoretical uncertainty in the knowledge of the antineutron-nucleus scattering length and also a smaller but nonzero uncertainty in the experimental knowledge of the neutron-nucleus scattering lengths, it will not be practical to tune the neutron-antineutron optical potential difference exactly to zero by this method. However one can reduce this energy difference enough that the neutron antineutron transition probability meets the so-called quasifree condition $\delta E T \leq \hbar$ where $T$ is the observation time in the experiment, and treat the remaining uncertainty in the cancellation accuracy as a systematic error in the experiment provided it is small enough. Note that one can measure easily the line integral of the magnetic field $\vec{B}$ along the neutron trajectories in a slow neutron beam $n-\overline{n}$ experiment by polarizing the neutrons and using the neutron spin rotation angle as a magnetometer, as was done successfully in the last free neutron experiment \cite{Baldo94} at the ILL.

A potential disadvantage of our proposed idea is that it works for one neutron polarization state. Many slow neutron polarizers either absorb or incoherently scatter one spin state out of the beam, thereby lowering the initial neutron intensity by at least a factor of 2.  Most of this loss can be avoided in principle through the use of a V-shaped transmission supermirror polarizer ~\cite{Mezei1977,Majkrzak1989}, which for slow neutrons can operate with better than 98\% efficiency and guides the neutron trajectories into two  different directions~\cite{Bazhenov1993}  and a spin flipper, which for slow neutrons can reach 99.9\% efficiency~\cite{Petukhov2019}. By flipping the slightly deflected neutrons, leaving the undeflected neutron unflipped, and redirecting the deflected neutrons back into the beam with mirrors, one can make the beam almost fully polarized  with losses much less than 50\%.

This paper shows that despite the common belief that the presence of gases and residual magnetic field suppress the probability of neutron-antineutron oscillations, the proper choice of the gas mixture and magnetic field can actually help a free neutron-antineutron oscillation experiment realize the free oscillation rate. In practice we expect that the main value of the possibility of operation of the experiment in this mode would be to relax some of the more difficult/expensive experimental conditions, especially the demand for a very small magnetic field. This requirement can be replaced with a larger magnetic field which requires less magnetic shielding material.  Of course the gas can only compensate a uniform magnetic field and not any variations. One still must ensure that the magnetic field does not vary significantly in magnitude~\cite{Davis2017}. The spatial variations in the magnetic field coming from
joints in the mumetal magnetic shield were already small enough in the ILL experiment \cite{Baldo94} to meet the quasifree condition. In the meantime great progress has been made in magnetic shielding technology developed for magnetically-shielded rooms \cite{Altarev:2014} and atom interferometry\cite{Wodey:2019ges} which has greatly reduced the amount of shielding needed to suppress nonuniformities and has developed a new understanding of how to treat the joints in the shields to more strongly suppress magnetic field leakage. One would need to employ this new knowledge to be able to take full advantage of our proposed operational mode.

The suggested approach can be used to operate a free $n-\bar{n}$ oscillation measurement in a magnetic field that is much larger than used in the previous ILL experiment \cite{Baldo94}. The ability to cancel the magnetic field and gas optical potentials is limited mainly by the accuracy of the theory calculation of the antineutron scattering lengths for hydrogen and helium. The neutron coherent scattering lengths of H$_{2}$, D$_{2}$ (a likely contaminant in the hydrogen gas), and $^{4}$He are all known to an absolute accuracy of better than $0.1$\%~\cite{Schoen2003,Haun2020} which is more than sufficient to help fix parameters in a theory calculation of the scattering lengths.
The gas density needed to fix the optical potential can be determined with high precision~\cite{Egan2019}. The absolute temperature of the gas can be determined to better than $0.1$\% near room temperature using platinum resistance thermometers, and the absolute pressure in the range of $1-100$ $\mu$bar of interest for this idea can be measured with an absolute precision of $0.1$\% using capacitive diaphragm gauges or spinning rotor pressure gauges. The optical potential of the gas is as uniform as the gas density, which in turn is set by the temperature uniformity of the gas volume, which can easily be controlled to be stable at better than the $0.1$\% level. The magnetic optical potential can likewise be determined with accuracy at or better than $0.1$\%. The neutron attenuation of parahydrogen gas at $100$ $\mu$bar pressure over 100 meters is below 1\% and therefore makes a negligible contribution to the decrease in neutron counting statistics.

For the off-diagonal components of the Hamiltonian which cause the oscillations,  there is an additional decoherence effect which places an upper bound on the gas density that can be employed in a $n-\bar{n}$ oscillation search.
As shown recently by Kerbikov~\cite{Kerbikov2018,Kerbikov2019}, the decoherent component of the neutron-matter interactions parametrized by the imaginary part of the neutron optical potential, which includes both neutron absorption and neutron incoherent scattering, can suppress the $n-\bar{n}$ oscillation rate. See~\cite{Harris1982} for a conceptually clear discussion of the suppression of oscillations of a two-state system due to decoherent interactions with an environment, which they use to explain the high stability of the handedness of chiral molecules, and others~\cite{Vacchini2008, Hemming2010} for earlier relevant two-state system calculations in agreement with these results.  This suppression of oscillations  can be calculated within the Lindblad formalism~\cite{Lindblad1976} for open quantum systems. This formalism was shown long ago~\cite{Lanz1997} to reproduce all of the usual results of the scattering theory for neutron optics, including the corrections to the ``usual'' scattering theory needed to satisfy the optical theorem~\cite{Sears}, and it shows clearly that the neutron absorption and incoherent scattering can all be viewed as sources of quantum decoherence.  For the  $n-\bar{n}$ system, the antineutron-nucleus absorption makes by far the dominant contribution to the suppression of oscillations from decoherence. The degree of suppression of the oscillations is determined by the absorption rate  $\lambda=n  v \sigma_{a}/2$ where $v$ is the neutron speed and $\sigma_{a}$ is the antineutron-nucleus absorption cross section. Kerbikov shows that, for the operating conditions of the ILL experiment and assuming that the residual gas was hydrogen, the probability to find an antineutron at time $t$ starting from an initial pure neutron state at time $t=0$ changes from~(\ref{prob})
\begin{equation}\label{Kerbikov}
P_{n \overline{n}}(t)=\frac{4{\delta m}^{2}}{\Omega^{2}}\exp \left[-(\lambda /2 + \Gamma )t\right]\sinh^2(\Omega t/2),
\end{equation}
where $\Omega^{2}={\lambda^{2}/4} - 4 {\delta m}^{2}$ and $\Gamma$ is the $n-\bar{n}$ beta decay rate. For the parameters considered in our example above the observation time  $t<< 1/\lambda$ is small compared to the inverse  absorption rate, so  this expression becomes

\begin{equation}\label{shorttime}
P_{n \overline{n}}(t)={\delta m}^{2} t^{2} -({1/2}){\delta m}^{2} \lambda t^{3},
\end{equation}

and this decoherent damping reduces the oscillation probability by less than 1\% compared to the undamped case. So although this form of decoherence is not a large effect for our proposed operational mode it would eventually become important at higher gas densities and/or longer observation times.  In the opposite limit $t>> 1/\lambda$ one can see from the equation above that the oscillation rate becomes exponentially suppressed.

\section{Acknowledgments}
This publication is funded by the Gordon and Betty Moore Foundation to support the work of V.V.N. and W.M.S. The authors are grateful to the participants of INT-17-69W Workshop "Neutron-antineutron oscillations: appearance, disappearance and baryogenesis" in Seattle, USA, held on 23-27 October 2017, as well as to our colleagues from the GRANIT collaboration. V.G. is grateful for support to the U.S. Department of Energy, Office of Science, Office of Nuclear Physics program under Award Number DE-SC0015882. W.M.S. acknowledges support from NSF PHY-1913789 and from the Indiana University Center for Spacetime Symmetries.

\bibliography{NNbar}

%\end{thebibliography}

\end{document}